\documentclass{aastex6}
\fullcollaborationName{Friends}

\begin{document}


\title{Consequences of a distant massive planet on the large semi-major axis Trans-Neptunian Objects}


\author{C. Shankman\altaffilmark{1}}
\affil{Department of Physics and Astronomy, University of Victoria, Elliott Building, 3800 Finnerty Rd, Victoria, British Columbia V8P 5C2, Canada}
\email{cshankm@uvic.ca}

\author{JJ. Kavelaars}
\affil{National Research Council of Canada, Victoria, BC, Canada}

\author{S. M. Lawler}
\affil{National Research Council of Canada, Victoria, BC, Canada}

\author{B. J. Gladman}
\affil{Department of Physics and Astronomy, The University of British Columbia, Vancouver, BC, Canada}

\and

\author{M. T. Bannister}
\affil{Astrophysics Research Centre, Queen’s University Belfast, Belfast BT7 1NN, United Kingdom}


\altaffiltext{1}{cshankm@uvic.ca}

\begin{abstract}
We explore the distant giant planet hypothesis by integrating the large semi-major axis, large pericenter Trans-Neptunian Objects (TNOs) in the presence of the giant planets and an external perturber whose orbit is consistent with the proposed distant, eccentric, and inclined giant planet, so called planet 9.
We find that TNOs with semi-major axes greater than 250 au experience some longitude of perihelion shepherding, but that a generic outcome of such evolutions is that the TNOs evolve to larger pericenter orbits, and commonly get raised to retrograde inclinations.
This pericenter and inclination evolution requires a massive disk of TNOs (tens of M$_\Earth$) in order to explain the detection of the known sample today.
Some of the highly inclined orbits produced by the examined perturbers will be inside of the orbital parameter space probed by prior surveys, implying a missing signature of the 9th planet scenario.
The distant giant planet scenarios explored in this work do not reproduce the observed signal of simultaneous clustering in argument of pericenter, longitude of the ascending node, and longitude of perihelion in the region of the known TNOs.

\end{abstract}




\section{Introduction} \label{sec:intro}

In Solar System studies, the observation of unexpected orbital behavior has been used to further our understanding of both physics and the composition of the Solar System.
Two exemplary successes of this process can be found in the prediction of the existence of Neptune from observing the evolving orbit of Uranus \citep{leverrier1846}, and the confirmation of General Relativity through its ability to explain the precession of Mercury's orbit \citep{leverrier1859,einstein1916}.
This pattern repeats itself in the outer Solar System today as we discover new mysteries in the Trans-Neptunian Object (TNO) population.

The decoupled perihelion of 2001 CR$_{105}$ \citep{gladmanetal02} demonstrated that some TNOs inhabit the domain that is dominated by the gravitational influence of Neptune. 
With a large semi-major axis $a$ of 226 astronomical units (au) and a pericenter $q$ of 44 au, well beyond the $q \lesssim 37$ au active influence of Neptune \citep{lykawkamukai07,gladman08}, 2001 CR$_{105}$ required a new dynamical path to emplace TNOs on orbits so separated from the influence of the giant planets.
\citet{gladmanetal02} explore various possible formation scenarios for 2001 CR$_{105}$ including a fossilized disk, a passing star, the presence of a small (lunar to Mars sized) planet beyond the orbit of Neptune, and others argue that 2001 CR$_{105}$ can be emplaced on its orbit by a complex path of planetary migration and resonance capture, but only for objects with $a < 260$ au \citep{gomes03,gomes05}.
The discovery of 2001 CR$_{105}$ was followed by surveys finding other TNOs with orbits outside the domain of Neptune (now numbering in the tens) that have revealed unexpected and difficult to explain structure in the outer Solar System. 

With the discovery of (90377) Sedna \citep{brownetal04}, an even more extreme orbit space was revealed.
Sedna, with its $q$ at 76 au and $a$ of 500 au, is not dynamically coupled to the giant planets nor to galactic tides. 
It is  difficult to form planetesimals at such a large distance given current models of planetesimal formation, and therefore the current orbits of these detached TNOs requires some dynamical interaction (ongoing or long past) to have emplaced them on such orbits.
Theories for emplacement of TNOs on large-$q$ orbits have included an additional planet in the Solar System \citep[e.g.][]{gladmanetal02,brownetal04,gomes06,soaresgomes13}, stellar flybys \citep[e.g.][]{idaetal00,kenyonbromley04,morbidellilevison04,kaib11b,soaresgomes13,brasserschwamb15} and ejected rogue planets \citep[e.g.][]{thommesetal02,gladmanchan06}.
These discoveries have continued, with one new extreme orbit TNO discovered every few years (e.g. 2004 VN$_{112}$, \citet{becker08}, and 2010 GB$_{174}$, \citet{chen13}).

In 2014, \citet{trujillosheppard14} reported the discovery of another large pericenter TNO, 2012 VP$_{113}$, with $q \simeq 80$.
\citet{trujillosheppard14} also noted a clustering (i.e. apparent grouping in the observed sample) of pericenter arguments $\omega$ near $0^{\circ}$ for all of the known TNOs having $a > 150$ au and $q > 30$ au.
There are known detection biases for discovering objects at their pericenters and in the ecliptic plane, which could explain an enhancement of detections of objects with $\omega = 180^{\circ}$ versus  $\omega = 0^{\circ}$, however there have been no TNOs detected with $\omega$ near $180^{\circ}$.
As of this writing, there has no published demonstrations that the lack of $\omega = 180^{\circ}$ TNOs results from an observation bias, nor has there been a demonstration that these detections should be free from such an observation bias.
\citet{trujillosheppard14} suggest that the clustering of TNOs with $\omega = 0^{\circ}$ might be explainable through a dynamical interaction with a massive perturber on a 250 au orbit (required to maintain the observed clustering over time); they posit the existence of a 9th planet of super-Earth-mass size in the outer Solar System as the cause of the observed $\omega$ clustering, but do not demonstrate a particular dynamical pathway to cause clustering around $\omega = 0$ lasting until the present day.


\begin{figure*}[h]
\centering
\includegraphics[scale=0.4]{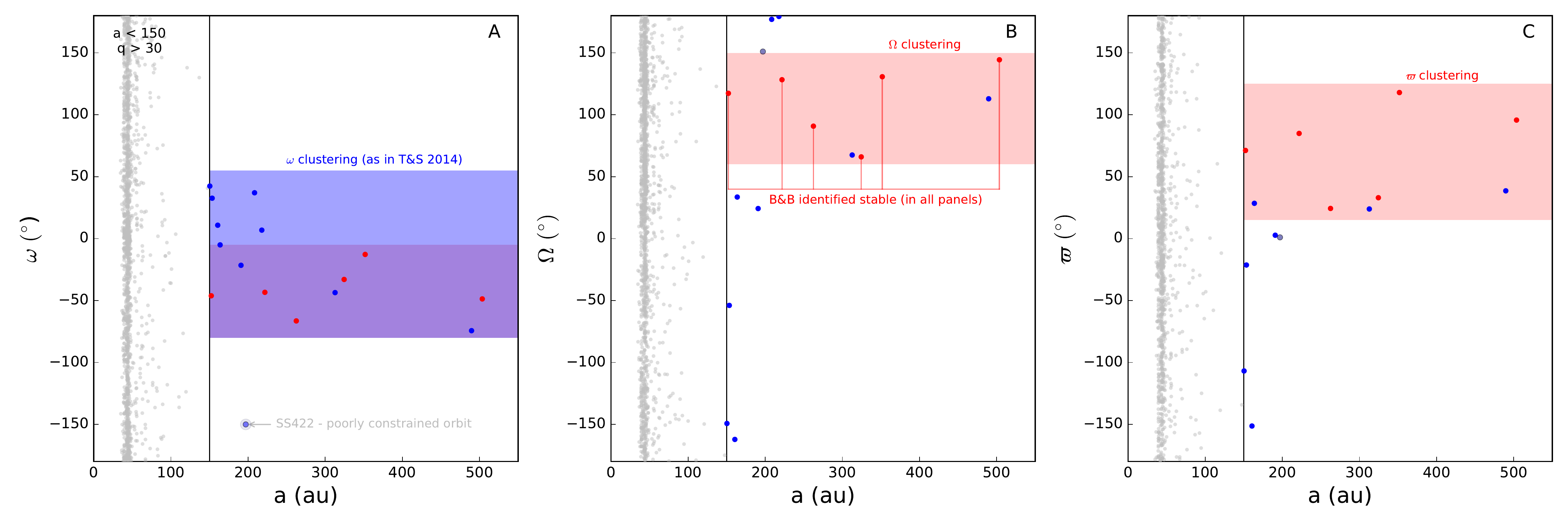}
\caption{
All TNOs with $q > 30$ from the Minor Planet Center (MPC) database as of 2016 May 3.
TNOs with $a < 150$ au are plotted in grey, the \citet{batyginbrown16} identified stable TNOs are plotted in red (Table~\ref{tab:TNOgroups}), and the remaining $a > 150$ au TNOs are plotted in blue (Table~\ref{tab:TNOlist}). 
Panel \textbf{A} plots argument of pericenter $\omega$ vs semi-major axis $a$, panel \textbf{B} plots longitude of the ascending node $\Omega$ vs semi-major axis $a$, and panel \textbf{C} plots longitude of perihelion $\varpi$ vs semi-major axis $a$.
Panel \textbf{A} highlights a clustering in $\omega$ as first reported in \citet{trujillosheppard14}, where panels \textbf{B} and \textbf{C} show $\Omega$ and $\varpi$ clustering as reported in \citet{batyginbrown16}.
The red and blue bands guide the eye to regions of clustering.
2003 SS$_{422}$ was not included in the analyses of \citet{trujillosheppard14} or \citet{batyginbrown16}, possibly due to its large $a$ uncertainty ($\Delta \, a \sim 50$ au); 2003 SS$_{422}$ is plotted in light blue in all panels and indicated in panel \textbf{A}.
}
\label{fig:angles}
\end{figure*}


In 2016, \citet{batyginbrown16} re-examined the $a > 150$ au TNOs, reporting that the large-$a$ TNOs that are not dynamically interacting with Neptune show clustering not only in $\omega$ but also in longitude of the ascending node $\Omega$ and thus also in longitude of pericenter, $\varpi = \omega + \Omega$ (see Figure~\ref{fig:angles}). 
A 9th planet (henceforth P9) with $a$ = 700 au, eccentricity $e$ = 0.6, inclination $i = 30^{\circ}$ and a mass of at least 10 Earth-masses \footnote{any distant planet must have a mass less than $M_{Saturn}$  to be consistent with a non-detection in the all-sky WISE survey\citep{luhman14}} is proposed as the cause of the observed confinement in $\varpi$.
They demonstrate analytically and with numerical simulations that their proposed P9 can produce $\varpi$ confinement for the age of the Solar System in test particles with $a > 250$ au and $\omega$ and $\Omega$ confinement for test particles with $a > 500$ au.
They also find that an inclined P9 produces highly inclined distant TNOs and propose that such a planet could explain the origin of the known large-$a$ TNOs with $i$ between 60$^{\circ}$ and $150^{\circ}$ (e.g. Drac 2008 KV$_{42}$, \citet{gladman09b}) whose formation was not securely identified \citep{elliot05,gladman09b, chen16}, but has been explained by a distant planet \citep{gomesetal15}. 
These arguments launched a flurry of discussions and studies on the origins, location and implications of a ninth planet. 
The studies have covered formation and capture scenarios \citep{bromleykenyon16,cowanetal16, kenyonbromley16,liadams16,mustilletal16}, constraints on the location, detectability and physical properties of P9 \citep{brownbatygin16,dlfmdlfm16b,dlfmdlfm16c,fiengaetal16,fortneyetal16,holmanpayne16,holmanpayne16b,lindermordasini16,philippov16,toth16}, the dynamical implications of P9 in the Solar System \citep{dlfmdlfm16a,lawleretal16}, resonances and P9 \citep{beust16,malhotraetal16}, a dark matter P9 \citep{sivarametal16}, and impacts of P9 on the Sun's obliquity \citep{baileyetal16,lai16}.

The studies to date have primarily focused on the formation and detection of P9, with few publications examining the impact on the observed TNO populations of a massive perturber in the distant Solar System.
\citet{lawleretal16} model the emplacement of the scattering and detached TNO populations in the presence of a super-earth planet beyond 200 au.
They find that such a planet sculpts distinctly different $e$ and $i$ distributions within the scattered and detached TNOs, but that this effect is not detectable in well-characterized, published surveys to date due to the strong flux bias at those distances.
\citet{dlfmdlfm16a} performed numerical integrations of the six known $a > 250$ au TNOs (see Table~\ref{tab:TNOgroups}) in the presence of the four giant planets and the \citet{batyginbrown16} P9. 
Integrating these TNOs for 200 Myr, \citet{dlfmdlfm16a} find that P9 may destabilize the orbits of several of these TNOs on timescales of dozens of Myr and can result in their ejection from the Solar System.
These two works are the first to study the implications of P9 on observed TNO populations, but to date there has been no work examining the implications on the detectability of the large-$a$ TNOs that were used to infer the existence of P9.

In this work we study the dynamical impact of P9 on the known large-$a$, large-$q$ TNOs (Table~\ref{tab:TNOlist}). 
Performing a set of n-body simulations, we examine the implications for the detectability of these TNOs and the evolution of their $\omega$, $\Omega$, and $\varpi$ angles to assess whether the proposed P9 reproduces the original observed clustering signal used to infer its existence.

\section{Methods} \label{sec:methods}

We test the P9 hypothesis by supposing the existence of such a planet in the Solar System today and exploring the implications for the large-$a$ TNOs. 
We perform a set of n-body integrations of the large-$a$ TNOs (see Figure~\ref{fig:angles}, Table~\ref{tab:TNOlist}) in the presence of the giant planets and a candidate P9, examining the implications of the presence of a massive perturber through the orbital evolution of the TNOs.
\citet{batyginbrown16} report possible $a$, $e$, $i$, $\omega$, and $\Omega$ values for P9 (see Table~\ref{tab:P9s}) and assume a mass of around 10 Earth-masses. 
The mean anomaly $\mathcal{M}$ of P9 is not well-constrained by observations.
A more distant P9 (i.e. $\mathcal{M}$ near 180$^\circ$) is more difficult to detect, and thus more consistent with the current non-detection. 
Resonance confinement is not the proposed mechanism for $\varpi$ shepherding between P9 and the large-$a$ TNOs and thus the choice of $\mathcal{M}$ does not affect the confinement; we therefore assign P9 $\mathcal{M} = 180^{\circ}$.
\citet{batyginbrown16} propose an $i = 30^{\circ}$ for P9, motivated by this configuration's ability to produce highly inclined TNOs.
We test the sensitivity of our analysis to this choice of inclination by also examining scenarios with P9 at lower inclinations of $15^{\circ}$ and $0^{\circ}$.
Table~\ref{tab:P9s} lists the orbital elements and mass for the three P9 configurations examined in this work; these parameters are consistent with current studies on the observable constraints \citep{brownbatygin16,fiengaetal16,holmanpayne16,holmanpayne16b}.
We also perform a control simulation without a P9.

\citet{trujillosheppard14} note an $\omega$-clustering for TNOs with $a > 150$ au and $q >30$ au and \citet{batyginbrown16} note that a subset of these ($a > 250$ au and $q > 30$ au) cluster in $\Omega$ and $\varpi$.
The choice of a semi-major axis boundary at $150$ au has not been given any physical motivation, however there is an apparent clustering of $\omega$ beyond this $a > 150$ au threshold (Figure~\ref{fig:angles} panel A) and so we continue this approach and select all of the TNOs satisfying $a > 150$ au and $q > 30$ au as the sample for this study\footnote{The sample was selected in May 2016.} for consistency and reproducibility with prior studies \citep{trujillosheppard14,batyginbrown16}.
Table~\ref{tab:TNOlist} lists this set of TNOs with their orbital elements and absolute magnitudes $H$ as reported by the MPC.

We perform n-body simulations with the MERCURY6 \citep{chambers99} suite.
The hybrid symplectic/Bulirsch-Stoer algorithm, which balances integration speed with modeling close encounters, was used with a base time step of 0.5 years \footnote{0.5 years is less than $\frac{1}{20} $th of Jupiter's orbital period that sets the shortest dynamical timescale in the simulation.} and all simulations were run for 4 Gyr. 
The giant planets and TNOs in Table~\ref{tab:TNOlist} were added to the simulation with orbital elements taken from the NASA Horizons database\footnote{http://ssd.jpl.nasa.gov/horizons.cgi} for the date of 2016 January 1 (JD 2457388.5). 
As N-body dynamics are inherently chaotic, we examine the evolution of each TNO through the evolution of a set of 60 clones created by sampling each object's orbit uncertainties. 
Orbit uncertainties for each TNO were taken from the JPL Small-body Database\footnote{http://ssd.jpl.nasa.gov/sbdb.cgi} (Table~\ref{tab:TNODelta}) and clones were generated in one of two ways: \textbf{a)} if the $a$ uncertainty was small ($\Delta \, a < 0.5$ au), 60 orbits within the $a, e, i, \omega, \Omega$ and true anomaly uncertainties were randomly and uniformly sampled; \textbf{b)} if $\Delta a > 0.5$ au, three clusters of orbits were generated - one cluster of 20 orbits at each of the $a, q$ extremes and one cluster of 20 orbits around the nominal orbit.
For clarity we emphasize that in our simulation all TNO and planet orbits and uncertainties are taken from the NASA Horizons database and not the MPC; Table~\ref{tab:TNOlist} lists the Minor Planet Center (MPC) elements and magnitudes as a convenient reference.
We test three cases of P9 inclinations (Table~\ref{tab:P9s}) and one control simulation with the 4 giant planets.
Each of our simulations thus includes 4 or 5 planets and 960 test particles; each was integrated for the 4 Gyr age of the Solar System.

\begin{deluxetable}{c * {8}{c}}
\tablecaption{All TNOs with $a > 150$ au and $q > 30$ au in the MPC database \label{tab:TNOlist}}
\tablehead{
\colhead{MPC} &  \colhead{$a$} &  \colhead{$e$} &  \colhead{$q$} &  \colhead{$i$}  &  \colhead{$\omega$} &  \colhead{$\Omega$} &  \colhead{$M$} &  \colhead{$H_v$}\\
\colhead{Designation} &  \colhead{(au)} & \colhead{}  &  \colhead{(au)} &  \colhead{($^{\circ}$)} &  \colhead{$(^{\circ})$} &  \colhead{$(^{\circ})$} &  \colhead{($^{\circ}$)} & \colhead{}
}
\startdata
Sedna &  499.08 &  0.85 &  76.04 &  11.9 &  311.5 &  144.5 &  358.1 &  1.6 \\ 
2007 TG$_{422}$ &  482.4 &  0.93 &  35.57 &  18.6 &  285.8 &  112.9 &  0.4 &  6.2 \\ 
2010 GB$_{174}$ &  369.73 &  0.87 &  48.76 &  21.5 &  347.8 &  130.6 &  3.3 &  6.5 \\ 
2013 RF$_{98}$ &  325.1 &  0.89 &  36.29 &  29.6 &  316.5 &  67.6 &  0.1 &  8.6 \\ 
2004 VN$_{112}$ &  317.71 &  0.85 &  47.32 &  25.6 &  327.1 &  66.0 &  0.4 &  6.4 \\ 
2012 VP$_{113}$ &  260.81 &  0.69 &  80.27 &  24.1 &  292.8 &  90.8 &  3.3 &  4.0 \\ 
2001 FP$_{185}$ &  226.86 &  0.85 &  34.26 &  30.8 &  7.0 &  179.3 &  1.3 &  6.2 \\ 
2000 CR$_{105}$ &  226.14 &  0.8 &  44.29 &  22.7 &  317.2 &  128.3 &  5.4 &  6.3 \\ 
2002 GB$_{32}$ &  217.9 &  0.84 &  35.34 &  14.2 &  37.0 &  177.0 &  0.3 &  7.8 \\ 
2003 SS$_{422}$ &  196.44 &  0.8 &  39.37 &  16.8 &  210.8 &  151.1 &  359.2 &  7.1 \\ 
2007 VJ$_{305}$ &  187.74 &  0.81 &  35.18 &  12.0 &  338.3 &  24.4 &  1.5 &  6.6 \\ 
2003 HB$_{57}$ &  165.36 &  0.77 &  38.1 &  15.5 &  10.9 &  197.8 &  1.3 &  7.4 \\ 
2015 SO$_{20}$ &  162.02 &  0.8 &  33.16 &  23.4 &  354.9 &  33.6 &  359.8 &  6.4 \\ 
2013 GP$_{136}$ &  153.33 &  0.73 &  41.11 &  33.5 &  42.2 &  210.7 &  356.2 &  6.6 \\ 
2010 VZ$_{98}$ &  151.89 &  0.77 &  34.32 &  4.5 &  313.9 &  117.4 &  357.8 &  5.1 \\ 
2005 RH$_{52}$ &  151.21 &  0.74 &  38.98 &  20.5 &  32.3 &  306.1 &  2.6 &  7.8 \\ 
\enddata
\tablecomments{Data downloaded on 2016 May 3. J2000 heliocentric orbital elements and absolute magnitudes are as reported by the MPC.}
\end{deluxetable}

\begin{deluxetable}{c * {8}{c}}
\tablecaption{Uncertainties in orbital parameters for all TNOs in Table~\ref{tab:TNOlist} \label{tab:TNODelta}}
\tablehead{
\colhead{MPC} &  \colhead{$\Delta$ $a$} &  \colhead{$\Delta$ $e$} &   \colhead{$\Delta$ $i$}  &  \colhead{$\Delta$ $\omega$} &  \colhead{$\Delta$ $\Omega$} \\
\colhead{Designation} &  \colhead{(au)} & \colhead{}  &  \colhead{($^{\circ}$)} &  \colhead{$(^{\circ})$} &  \colhead{$(^{\circ})$}  
}
\startdata
Sedna           &  0.78  &  2.39e-4 &       3.0416e-05 &       1.34e-2  &        1.60e-3   \\
2007 TG$_{422}$ &  3.45  &  5.24e-4 &       2.45e-4 &       2.93e-2  &        1.74e-3   \\
2010 GB$_{174}$ &  28.30  &  1.10e-2   &       5.42e-3  &       3.79e-1   &        2.09e-2    \\
2013 RF$_{98}$  &  36.67  &  1.46e-2   &       7.77e-2   &       5.54    &        1.26e-1      \\
2004 VN$_{112}$ &  1.07   &  4.85e-4 &       3.69e-4 &       1.04e-2  &        6.79e-4  \\
2012 VP$_{113}$ &  7.13  &  1.08e-2   &       4.15e-3  &       2.40    &        9.96e-3   \\
2001 FP$_{185}$ &  0.32 &  2.10e-4 &       4.74e-4 &       1.19e-2  &        1.05e-4  \\
2000 CR$_{105}$ &  0.53 &  4.54e-4 &      4.01e-4 &       9.88e-3 &        2.16e-4  \\
2002 GB$_{32}$  &  0.68 &  4.92e-4 &       2.27e-4  &       3.82e-3 &        3.53e-4  \\
2003 SS$_{422}$ &  47.83  &  5.56e-2   &   5.89e-2   &       17.18    &        6.98e-2    \\
2007 VJ$_{305}$ &  0.55 &  5.25e-4 &       8.47e-4 &       4.58e-2  &        1.31e-3   \\
2003 HB$_{57}$  &  0.58 &  7.73e-4 &       9.66e-4 &       5.48e-2  &       3.96e-4  \\
2015 SO$_{20}$  &  0.13 &  1.62e-4 &       6.24e-4 &       2.18e-2  &        2.84e-4  \\
2013 GP$_{136}$ &  0.57 &  1.17e-3  &       1.30e-3  &       1.14e-1    &    1.61e-4  \\
2010 VZ$_{98}$  &  0.15 &  2.10e-4 &       6.11e-05 &       7.58e-3 &        3.57e-3    \\
2005 RH$_{52}$  &  0.19 &  2.91e-4  &      5.90e-4 &       6.84e-2  &        1.48e-3   \\
\enddata
\tablecomments{1-$\sigma$ uncertainties taken from the JPL small-body database(http://ssd.jpl.nasa.gov/sbdb.cgi). All values are J2000 heliocentric, generated for the JD  2457600.5 (2016 Jul 31)}
\end{deluxetable}

\begin{deluxetable}{c * {8}{c}}
\tablecaption{Examined TNO groups \label{tab:TNOgroups}}
\tablehead{
\colhead{Category} & \multicolumn{6}{c}{Designation}
}
\startdata
$a > 250$ au & Sedna & 2010 GB$_{174}$  & 2004 VN$_{112}$ & 2012 VP$_{113}$ & 2007 TG$_{422}$ & 2013 RF$_{98}$ \\
\citet{batyginbrown16} identified stable & Sedna & 2010 GB$_{174}$ & 2004 VN$_{112}$ & 2012 VP$_{113}$ & 2000 CR$_{105}$ & 2010 VZ$_{98}$ \\
\enddata
\end{deluxetable}

\begin{deluxetable}{c * {8}{c}}
\tablecaption{P9 configurations tested. \label{tab:P9s}}
\tablehead{
\colhead{P9} &  \colhead{$a$} &  \colhead{$e$} & \colhead{$q$} &  \colhead{$i$} &  \colhead{$\omega$} &  \colhead{$\Omega$} &  \colhead{$M$} & \colhead{Mass} \\ 
\colhead{} &  \colhead{(au)} & \colhead{}  &  \colhead{(au)} &  \colhead{($^{\circ}$)} &  \colhead{$(^{\circ})$} &  \colhead{$(^{\circ})$} &   \colhead{($^{\circ}$)} & \colhead{}  
}
\startdata
P9 \citep{batyginbrown16} & 700 &  0.6 & 280 &  30 &  150 &  100 &  180 & 10 M$_\Earth$ \\
P9 (moderate $i$) & 700 &  0.6 & 280 &  15 &  150 &  100 &  180 & 10 M$_\Earth$ \\
P9 (low $i$) & 700 &  0.6 & 280 &  0 &  - &  - &  180 & 10 M$_\Earth$ \\
\enddata
\tablecomments{With an inclination of 0$^{\circ}$, $\omega$ and $\Omega$ become undefined, but $\varpi$ remains 150$^\circ$.}
\end{deluxetable}


\section{Results}
\label{sec:results}


\subsection{Signal in Angle Clustering}
\label{sec:results1}

P9 was proposed to explain the apparent clustering of one or more of the angles that determine an orbit's orientation.
Here we examine the shepherding of $\omega$, $\Omega$, and $\varpi$ through our dynamical simulations of the large-$a$ TNOs.
We focus our discussion of the results on the nominal P9 case with an inclination of 30$^{\circ}$, checking that our results hold for different choices of P9 inclination.

\begin{figure*}[h]
\centering
\includegraphics[width=\textwidth,height=\textheight,keepaspectratio]{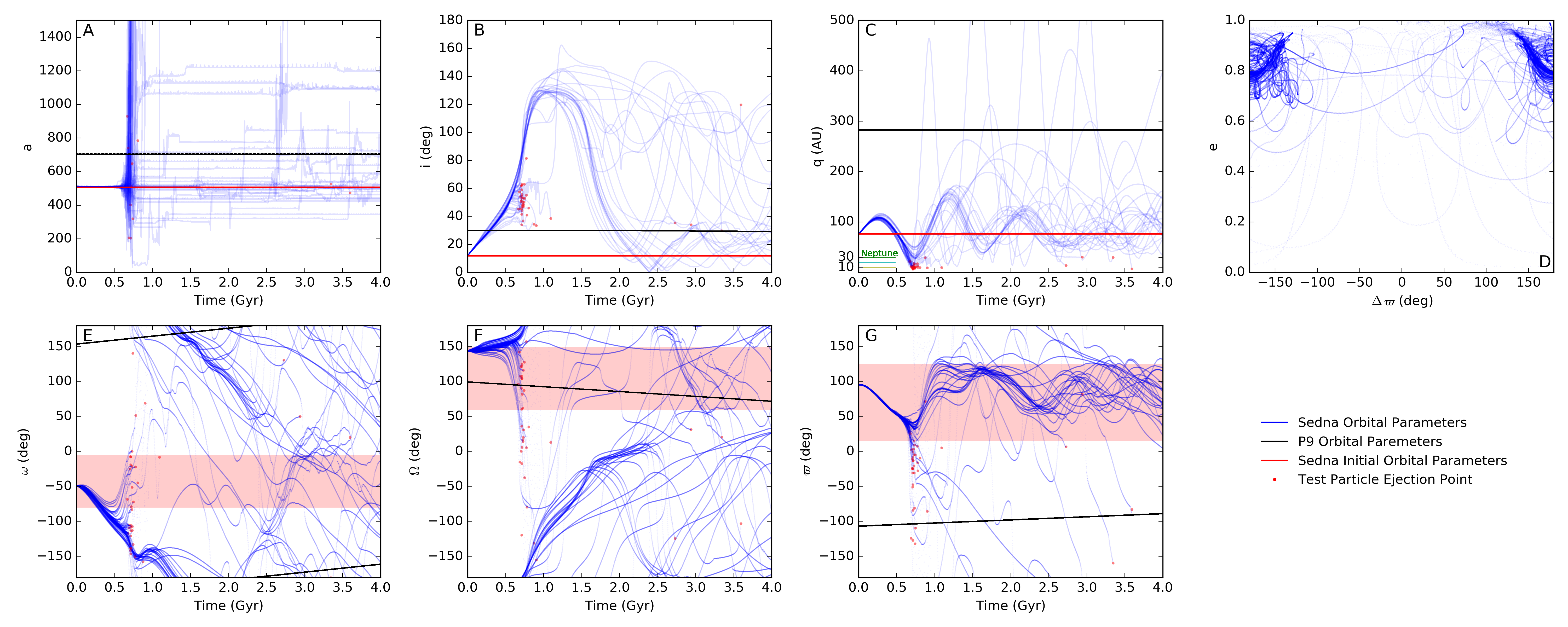}
\caption{
The orbital evolution for the barycentric orbital elements of the 60 clones of Sedna across the 4 Gyr simulation. 
The time evolution of clones are plotted in blue in all panels. 
Panel \textbf{D} plots the $\varpi_{\textrm{P9}} - \varpi_{\textrm{clone}}$ vs $e$, which shows the secular interaction between the clones and P9. 
The red horizontal line in panels \textbf{A}, \textbf{B}, and \textbf{C} mark the present-day observed values for Sedna's orbital elements plotted in those panels. 
The black line plots the orbital evolution of P9. 
Red circles mark the ejection point of a clone from the simulation ($a > 10 000$ au or collision with the Sun). 
The red bands in panels \textbf{E}, \textbf{F}, and \textbf{G} mark the region of confinement in the real TNOs for $\omega$, $\Omega$, and $\varpi$, respectively (as in Figure~\ref{fig:angles}).
The orbits of the giant planets are plotted in Panel \textbf{C}, showing when clones are driven into giant planet crossing orbits.
}
\label{fig:VB12}
\end{figure*}

\begin{figure*}[h]
\centering
\includegraphics[width=\textwidth,height=\textheight,keepaspectratio]{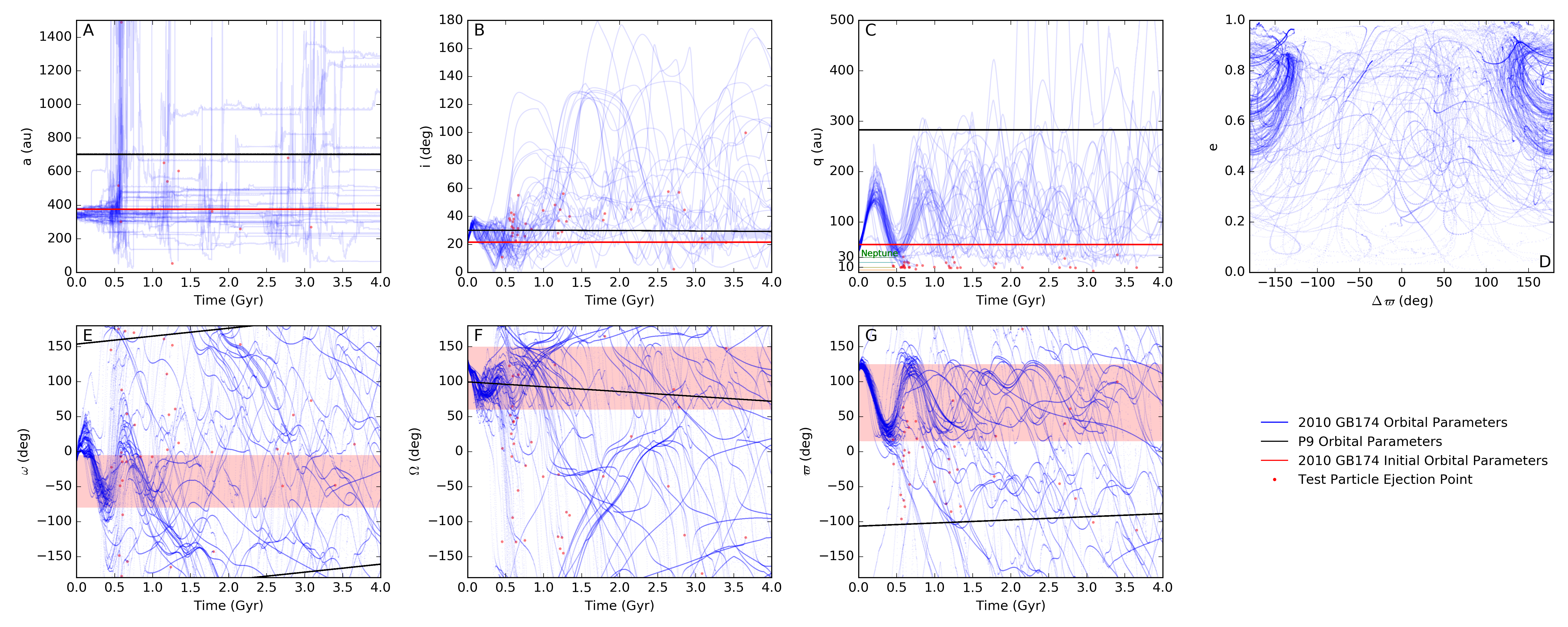}
\caption{
The orbital evolution for the 60 clones of 2010 GB$_{174}$ across the 4 Gyr simulation. 
Line styles and panels are as described in Figure~\ref{fig:VB12}.
}
\label{fig:GB174}
\end{figure*}

Here we examine the evolution of two of the TNOs in the sample as examples of the evolutions seen in our simulations\footnote{For plots of all TNOs across all simulations, see https://github.com/cshankm/P9-simulation-plots/tree/master/i30 [DOI to be made for time of publication]}.
Figure~\ref{fig:VB12} shows the orbital evolution for all of the clones of Sedna. 
The majority of Sedna clones undergo $\varpi$ \textit{shepherding} (i.e. driven confinement in a band) throughout the 4 Gyr simulation (Figure~\ref{fig:VB12} panel G), remaining in the region of the detected sample (red band).
The orbital evolution of 2010 GB$_{174}$ is plotted in Figure~\ref{fig:GB174}, showing that 2010 GB$_{174}$ undergoes similar $\varpi$ shepherding.
Of the 16 TNOs in the sample (Table~\ref{tab:TNOlist}), only six TNOs undergo $\varpi$ shepherding.
These confinements last for periods ranging from hundreds of Myr to 4 Gyr (e.g. see Figures~\ref{fig:VB12} and \ref{fig:GB174} panel \textbf{G}).
We find that only TNOs with $a > 250$ au (see Table~\ref{tab:TNOgroups}) undergo $\varpi$ shepherding. 
The analysis of \citet{batyginbrown16} highlights both the set of $a >250$ au TNOs and six stable TNOs (see Table~\ref{tab:TNOgroups}). 
We find that for TNOs with $q$ beyond Neptune, a large semi-major axis is the determining factor in $\varpi$ shepherding; all of the clones of the $a > 250$ au TNOs experience shepherding, however we find that two of the \citet{batyginbrown16} identified stable TNOs do not undergo this shepherding.
A massive, eccentric, external perturber generically drives $q$ down to Neptune coupling (Figures~\ref{fig:VB12} and \ref{fig:GB174} panel C), which then shuts off $\varpi$ shepherding.

As a necessary implicit condition of $\varpi$ shepherding, P9 also drives the evolution of $\omega$ and $\Omega$ (Figures~\ref{fig:VB12} and \ref{fig:GB174} panels E and F). 
Clones for all six of the $a > 250$ au TNOs (Table~\ref{tab:TNOgroups}) undergo $\varpi$ shepherding, and thus experience correlated $\omega$ and $\Omega$ evolution, but only two (VP$_{113}$ \& GB$_{174}$) experience shepherding of $\omega$ and $\Omega$, while the rest drift in a pattern that is mostly consistent with long-term circulation at a rate unique to each TNO.
Whether their $\omega$ and $\Omega$ are shepherded or circulating, the clones experience $\varpi$ confinement in the same region.

\begin{figure*}[]
\centering
\includegraphics[width=\textwidth,height=\textheight,keepaspectratio]{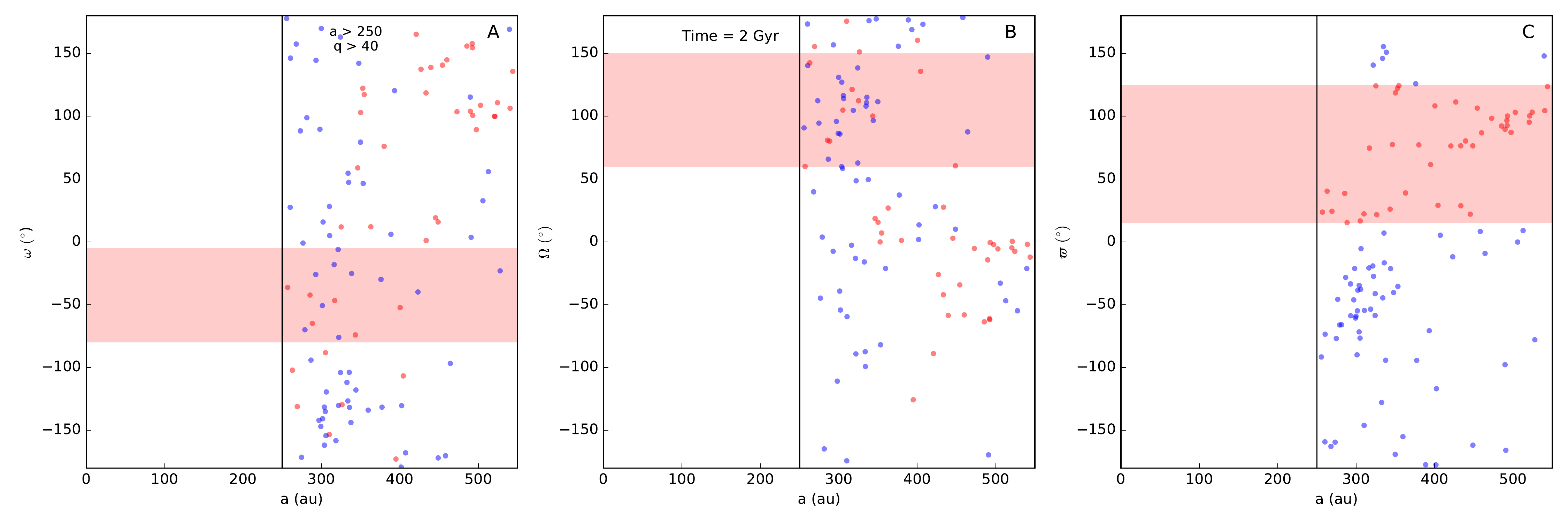}
\caption{$\omega$, $\Omega$, and $\varpi$ vs $a$ for all clones not strongly interacting with Neptune ($q >$ 40) and with a semi-major axis in the region of $\varpi$ shepherding ($a > 250$ au) at 2 Gyr into the simulation (choice of epoch does not affect the result).
Clones with $\varpi$ values that are in the observed TNO $\varpi$ clustering band are colored red in all panels, otherwise clones are plotted in blue.
Panels \textbf{A}, \textbf{B}, and \textbf{C} demonstrate that P9 does not sculpt $\omega$, $\Omega$ nor $\varpi$ in a restricted range as seen in the observed sample (Figure~\ref{fig:angles}).
Panels \textbf{A} and \textbf{B} demonstrate that for $\varpi$ shepherded clones, P9 does not simultaneously shepherd $\omega$ and $\Omega$ - this is also visible by comparison of panels E and F in Figures \ref{fig:VB12} and \ref{fig:GB174}.
}
\label{fig:angle_confinement}
\end{figure*}

Figure~\ref{fig:angle_confinement} explores the implication of this evolution on the initial signal used to infer the existence of P9: simultaneous $\omega$, $\Omega$, and $\varpi$ clustering.
The clones in Figure~\ref{fig:angle_confinement} are color coded red if they are in the region of $\varpi$ confinement at 2 Gyr \footnote{2 Gyrs is taken as a representative snapshot for visualization, but the results does not depend on epoch choice.} into the simulation (panel C) and blue otherwise.
Panels A and B show that clones in the $\varpi$ confinement band can occupy all values in $\omega$ and $\Omega$.
This is consistent with the result that the explored distant massive planets do not shepherd $\omega$ nor $\Omega$ in the region of the observed TNOs \citep{batyginbrown16,lawleretal16}. 
Panel C shows that while some clones experience $\varpi$ shepherding (Figures \ref{fig:VB12} and \ref{fig:GB174})), the influence of P9 does not sculpt a dominant band of confined $\varpi$ values.
This implies that there should be a large number of detectable TNOs at all values of $\omega$, $\Omega$ and $\varpi$.

\newpage


\newpage

\subsection{Orbital Evolution}
\label{sec:results2}

A 10 M$_\Earth$ planet with its $q$ at 280 au has a significant effect on the orbital evolution of the TNOs in our sample. 
In their explorations of the $a > 250$ au TNOs, \citet{dlfmdlfm16a} demonstrate that P9 can destabilize the orbit of these TNOs on short timescales ($< 200$ Myr); we extend the sample to include all TNOs with $ a> 150$ au and $q > 30$ au, and we also extend simulations to the age of the Solar System, revealing important structure in the $i$ and $q$ evolution of these TNOs.

\subsubsection{Perihelia Cycling}

Gravitational interactions with P9 raise and lower the pericentres of all of the clones in the sample, lowering some clones down into Neptune, or even Jupiter, crossing orbits or raising them out to hundreds of au (e.g. see Figures~\ref{fig:VB12},~\ref{fig:GB174}). 
All of the $\varpi$-shepherded clones undergo $q$ oscillations, which occur on roughly the same timescale as their $\varpi$ oscillations (see Figures~\ref{fig:VB12} and \ref{fig:GB174} panel C).
With $q$ cycled between hundreds of au and tens of au, the idea of a gap in overall distribution of TNO perihelia in the 50 - 70 au range \citep{trujillosheppard14} is incompatible with the P9 hypothesis.
The cycling $q$ affects the stability of the clones as they are pushed into planet-crossing orbits, and thus affects their survivability on Gyr timescales.
This increases the likelihood of ejection for the clones of TNOs that already currently interact with Neptune, and introduces instability for those which are presently on Neptune decoupled orbits (e.g. Sedna, 2012 VP113). 
Table~\ref{tab:ejection} gives the fraction of clones ejected for each TNO in our sample across the four simulations performed.
In contrast, the cycling to high values reduces the detectability of the clones.
This then requires a larger than previously expected reservoir of large-$a$ TNOs to explain the detection of the presently observed sample.

\clearpage

\begin{deluxetable}{c * {4}{c}}
\tablecaption{Fraction of clones ejected over the age of the Solar System for the different P9 inclinations tested. \label{tab:ejection}}
\tablehead{
\colhead{} & \multicolumn{4}{c}{\% Ejected} \\
\colhead{Designation} & \colhead{30$^{\circ}$} & \colhead{15$^{\circ}$} & \colhead{0$^{\circ}$} & \colhead{no P9} 
}
\startdata
Sedna        &   63   &   37  &    7   &    0 \\
2007 TG$_{422}$   &   67   &   95  &    82  &    88 \\
2010 GB$_{174}$   &   63   &   60  &    53  &    8 \\
2013 RF$_{98}$    &   93   &   93  &    90  &    52 \\
2004 VN$_{112}$   &   47   &   62  &    62  &    0 \\
2012 VP$_{113}$   &   77   &   52  &    75  &    0 \\
2001 FP$_{185}$   &   90   &   82  &    78  &    63 \\
2000 CR$_{105}$   &   75   &   63  &    53  &    0 \\
2002 GB$_{32}$    &   88   &   85  &    83  &    82 \\
2003 SS$_{422}$   &   82   &   83  &    77  &    42 \\
2007 VJ$_{305}$   &   82   &   88  &    88  &    73 \\
2003 HB$_{57}$    &   70   &   62  &    65  &    28 \\
2015 SO$_{20}$    &   75   &   82  &    85  &    88 \\
2013 GP$_{136}$   &   33   &   25  &    10  &    10 \\
2010 VZ$_{98}$    &   68   &   93  &    90  &    93 \\
2005 RH$_{52}$    &   63   &   40  &    38  &    17 \\
\enddata
\tablecomments{The clones sample the range around plausible orbits for each TNO, but are not constructed to represent the possible orbits given the ephemerides. For this reason, this table should not be interpreted as giving the stability for the actual orbit of the TNOs in the sample.}
\end{deluxetable}


\subsubsection{Sedna Population Estimate}

We examine the implications for the detection of Sedna (the most massive of the known $a > 150$ au TNOs) in detail as an example of the effects P9 induces on $q$ and $i$ evolution on the large-$a$ TNOs. 
Of the 60 clones of Sedna, 63\% of the clones are ejected during the simulation (see Table~\ref{tab:ejection}).  
For the clones of Sedna that were not ejected, each spent on average $\sim$~45\% of the 4 Gyr simulation with $q$ beyond the limit of detectability (of Sedna's discovery survey) due to $q$ raising, implying a large population of undetectable large-$a$ Sedna-like TNOs. 
As Sedna clones spend on average $\sim$~45\% of the simulation with $q$ beyond the detection limit, roughly one half of the population must be on completely undetectable orbits today.

The shape of an orbit, which sets the fraction of the orbit inside the detectable volume of surveys, affects the expected number of TNOs in the population that are required to explain the detected sample.
The detection of a TNO like Sedna which is on such large-$a$ orbit, where most of the orbit lies beyond a detection threshold, implies that there must be many TNOs on similar orbits in order for it to be probable to have detected Sedna near its pericenter.
Estimates of the size of these populations are regularly computed by asking what fraction of the time each TNO spends in an observable part of its orbit, which then gives the number of TNOs required for the expected detection of one object (as was done for Sedna in \citet{brownetal04,schwambetal09}).
We examine all of the clones of Sedna, at all time points in the simulation, and compute the population estimate required for the detection of those clones that have $q$ within the detection threshold.
While detectable, the Sedna clones require a population of TNOs on their orbit ranging from tens to, at times, the high thousands. 
We find a mean population estimate of $\sim$~80 Sedna-sized objects (across all clones and all time steps) are required for the detection of Sedna.
This estimate is double the best estimate for the Sedna population \citep{schwambetal09} but within the 1~$\sigma$ upper limit for their size of the Sedna population.
We combine our mean population estimate (80) with our findings that less than half of the clones of Sedna survive, and that surviving clones are only visible during half of the 4 Gyr simulation.
In the P9 scenario, the single detection of Sedna today requires a mass of 6-24 M$_\Earth$ from the ensemble of Sedna-like TNOs (down to absolute magnitude\footnote{The absolute magnitude, $H$, distribution transitions to a different form faint of $H_r$ = 8, which is diameter $D$ = 60 km for 16\% albedo.
A single slope absolute magnitude distribution with a slopes of 0.8 and 0.9 were used to determine the range of mass estimates. 
These slopes are consistent with the measurements of other TNO populations and the TAOS limit on the Sedna population \citep{wangetal09} 
An albedo of 16\% and density of 1 g/cm$^{3}$ were used for the mass estimates.} $H_r$ = 8).
This is more then an order of magnitude greater than the mass estimate required for the case of no P9.

\subsubsection{Inclination Raising}

\begin{figure*}[h]
\centering
\includegraphics[width=0.5\textwidth,height=0.5\textheight,keepaspectratio]{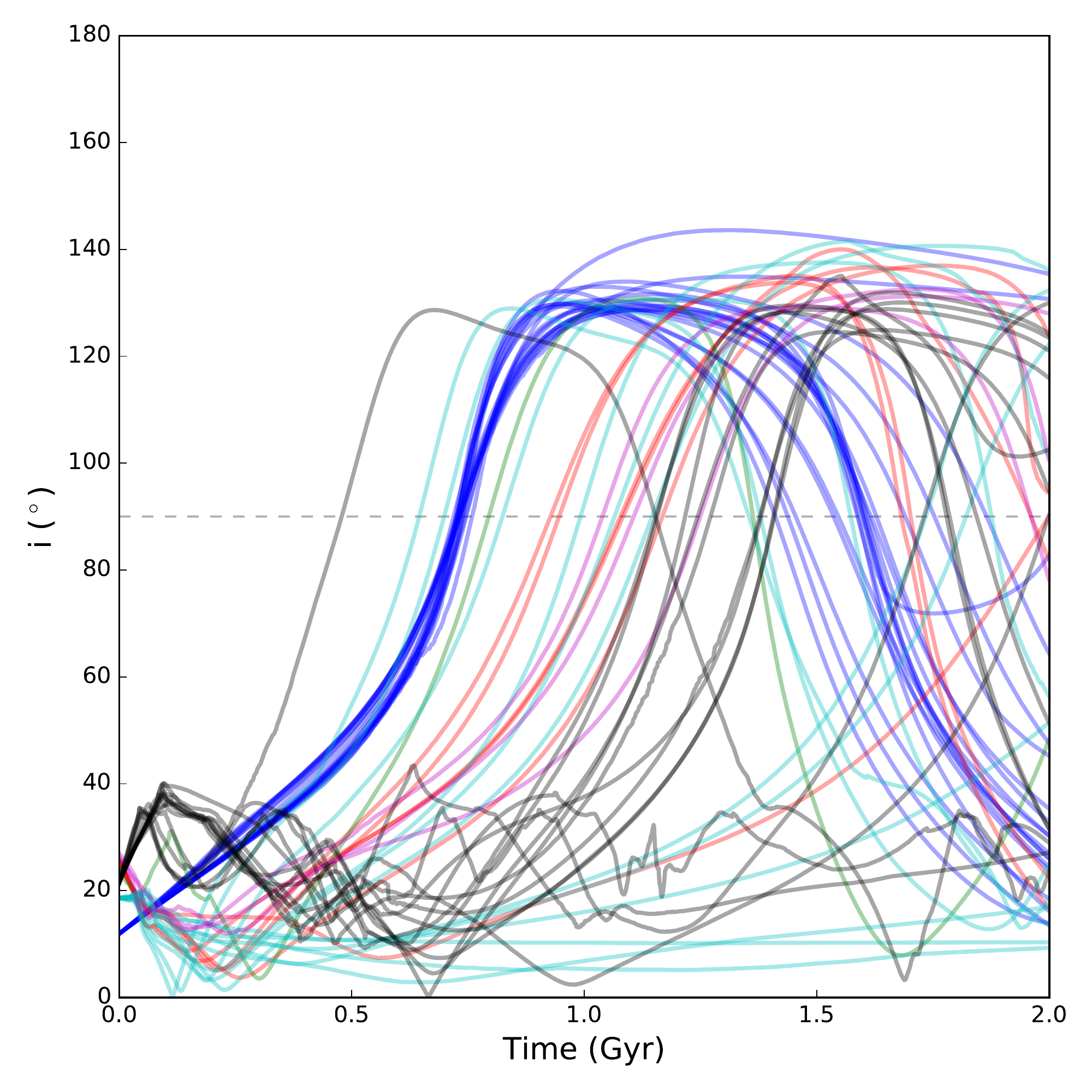}
\caption{
The $i$-evolution for all clones in our simulation that undergo $\varpi$ shepherding for the first 2 Gyr. 
Each color represents clones from one of the six $a > 250$ au TNOs (Table~\ref{tab:TNOgroups}). 
A dashed black line marks $i = 90^\circ$.
The color coding serves to demonstrate that clones of each TNO undergo the same generic evolution of $i$ cycling, which goes through extreme and retrograde values.
Raising inclinations through retrograde values is a generic feature of the dynamical mechanism that causes $\varpi$ shepherding.
}
\label{fig:i_confinement}
\end{figure*}

Perturbations from P9 raise and lower the inclinations of all TNOs in the sample. 
Figures~\ref{fig:VB12} and ~\ref{fig:GB174} panel B show that, on Gyr timescales, P9 raises the inclination, even into the retrograde state, before cycling them back to lower inclination. 
Examining only the clones that undergo $\varpi$ shepherding, we find that this $i$ flipping is a characteristic feature from the secular interactions (see Figure~\ref{fig:i_confinement}).
The P9 hypothesis implies the existence of an undetected, but potentially detectable, population of high-$i$ and retrograde large-$a$ TNOs.
Clones spend over half of the time with $i > 30^\circ$ suggesting that for each detected TNO there should be at least one undetected TNO at a higher inclination.
This implies that in a P9 Solar System, the mass estimates above should be factors of several larger.

\begin{figure*}[h]
\centering
\includegraphics[width=\textwidth,height=\textheight,keepaspectratio]{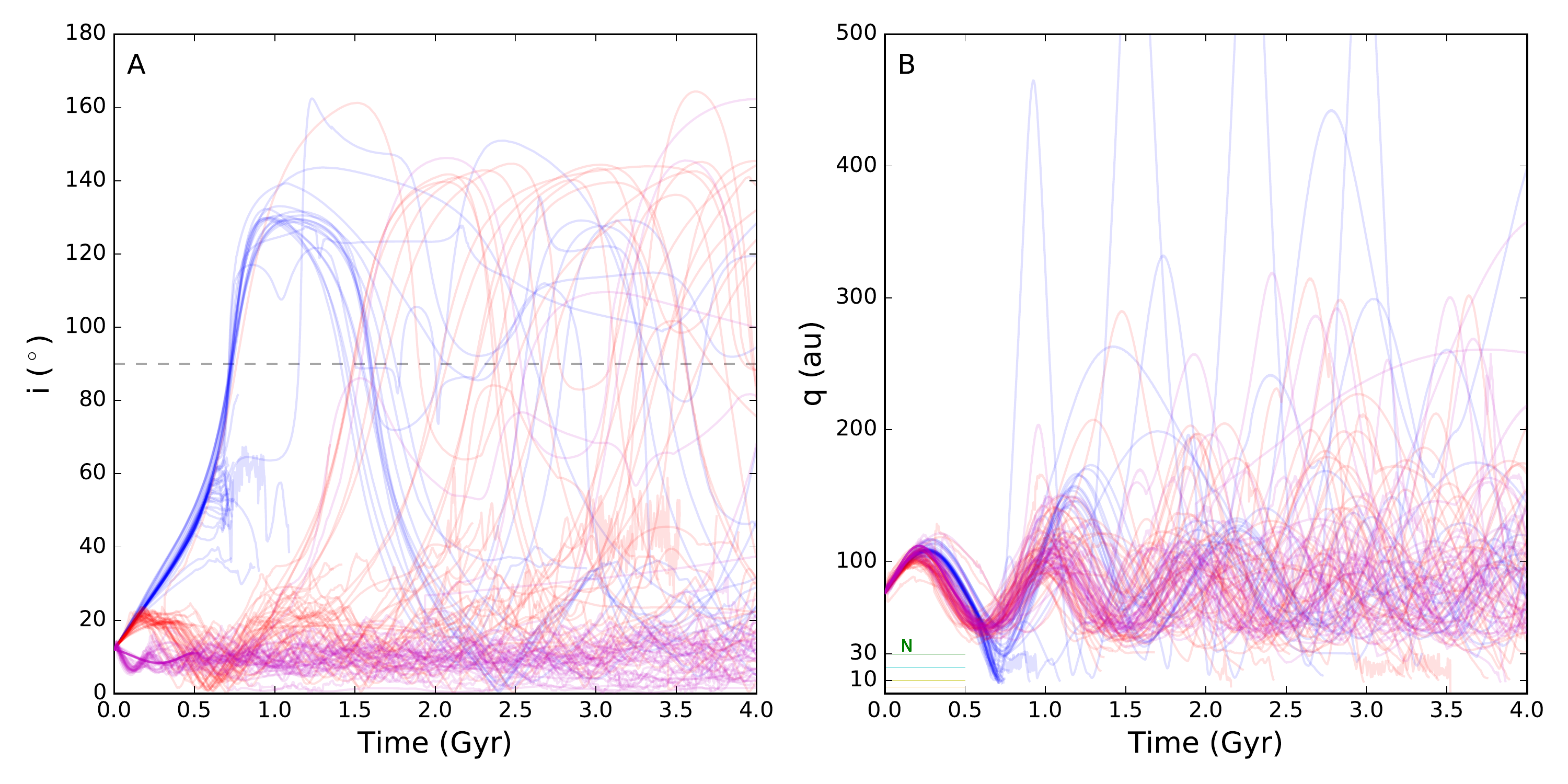}
\caption{
The $i$, $q$ evolution of the Sedna clones for the various P9 configurations. 
Clones from the simulation with $i = 30^\circ$ are shown in blue, P9 $i = 15^\circ$ in red and P9 $i = 0^\circ$ in magenta.
A dashed black line marks $i = 90^\circ$ in Panel \textbf{A}.
Panel \textbf{A} shows that the same $i$ raising and flipping is induced for $\varpi$ shepherded clones in both the P9 $i = 30^\circ$ and P9 $i=15^\circ$ cases.
Panel \textbf{B} shows that large $q$ oscillations occur for $\varpi$-shepherded clones in all simulations.
The cycling of $q$ is a generic feature of a massive external perturber, regardless of the perturber's inclination.
Inclination raising occurs generically for $\varpi$ shepherded clones in the simulations with high ($30^\circ$) and more moderate ($15^\circ$) perturber inclinations.
}
\label{fig:iqSedna}
\end{figure*}

We test the sensitivity of our results to the choice of P9 inclination.
In all simulations, we find that $\varpi$ shepherding is only induced for $a > 250$ au clones.
Figure~\ref{fig:iqSedna} panel B shows that the $q$ oscillations occur across all choices of P9 $i$ for $\varpi$ shepherded clones.
Inclination raising and flipping occurs for both P9 simulations with a non-zero $i$ perturber and takes the same form of raising inclinations through extreme and retrograde values that then cycle back to small values (Figure~\ref{fig:iqSedna} panel A).
The results in this work are independent of the choice of P9 inclination, with the exception that a zero degree inclination for P9 does not induce the same raising and flipping of inclinations in the clones of $\varpi$ shepherded TNOs.


\newpage

\section{Summary and Discussion}
\label{sec:discussion}

\textbf{1. } 
The P9 scenarios explored do not generically induce $\omega$ shepherding for the region of the observed TNOs ($a < 500$ au), a fact already noted in \citet{batyginbrown16}. 
This holds true even for TNOs that exhibit $\varpi$ shepherding.
The apparent clustering of $\omega$ is the initial signal that motivated the current incarnation of the hypothesis of a distant planet in the Solar System. 
The P9 scenarios explored in this work do not reproduce the observed signal of simultaneous clustering in $\omega$, $\Omega$, and $\varpi$ in the region of the detections (Figure~\ref{fig:angles}).

\textbf{2. } 
Clones that undergo $\varpi$ shepherding have their inclinations lifted to retrograde values. 
If there is a massive distant 9th planet, then there should be a significant number of lower-$q$, large-$a$, large-$i$ (or even retrograde) TNOs; only two retrograde TNOs (2008 KV$_{42}$ or Drac, \citet{gladman09b}, and 2011 KT$_{19}$ or Niku, \citet{chen16}) are known today, both of which have small semi-major axes.
Additionally, clones of TNOs that undergo $\varpi$ shepherding (i.e. those with $a > 250$ au) spend a significant portion of the age of the solar system with large inclinations, but all of the large-$a$ TNOs were detected with $i < 30^{\circ}$.
While there is a bias towards detection at low-$i$, Sedna was detected in an all-sky survey with sensitivity to high-$i$ and other surveys have looked at high latitudes (e.g. \citet{schwambetal09},  NGVS \citet{chen13}, CFEPS \citet{petit16})  but none have found any large-$a$ ($> 100$ au), high-$i$ TNOs.
The lack of detections of highly inclined TNOs poses a challenge for the inclined P9 scenario.

\textbf{3. }
During their orbital evolutions, $\varpi$-shepherded clones evolve through $q > 80$ au orbits where they would not be detected.  
The existence of a few $q\sim 40 - 80$ au detections implies a very massive (tens of M$_{\Earth}$) distant-$q$ reservoir and thus an efficient mechanism for delivery of material into this distant zone of the Solar System.
Implanting $\gtrsim 10 \textrm{M}_{\Earth}$ of solids on large-$a$, large-$q$ orbits with a high efficiency in the range of 10\% to 1\% would requires an implausibly large initial planetesimal disk with, respectively, hundreds to thousands of M$_{\Earth}$ of solids.

\textbf{4. } 
The existence of a distant massive planet destabilizes the orbits of the large-$a$ TNOs, like 2000 CR$_{105}$, 2004 VN$_{112}$, and 2012 VP$_{113}$, that would otherwise remain stable for the age of the solar system.  
Such a scenario thus requires an active supply reservoir and mechanism to deliver objects onto these orbits.
The perturber itself may act as the supply for this region, but this avenue was not explorable with these simulations.
The pericenter raising of a distant massive perturber will necessarily populate orbits with $q$ between 50 - 70 au, and therefore the scenario is inconsistent with the suggestion of a gap in the $q$ distribution proposed in \citet{trujillosheppard14}.

\textbf{5. }
As has been recently noted by \citet{sheppardtrujillo16}, some of the observed clustering in $\Omega$ may, in fact, be the result of observing bias.
Given that the observed TNOs cluster in $\omega$ near 0, and given a strong bias to detecting the TNO at pericenter, the $\Omega$ detected is determined by the direction of the survey pointing.
The location of survey pointings is determined by galactic plane avoidance and local weather conditions that are season dependent.
These biases may be able to explain the apparent clustering in $\Omega$ of the detected sample; future work is required to examine the effects of this bias in detail.
While this manuscript was in preparation, \citet{sheppardtrujillo16} report the discovery of several new large-$a$ TNOs that would fall within the sample explored here. 
Two of these TNOs have $\Omega$ values that broaden the range in the observed sample, which suggests that this apparent clustering will be eroded with future detections.

A bias in $\Omega$, given the clustering in $\omega$, would cause apparent $\varpi$ clustering. 
In order to explain the observed sample, the successful theory must be able to explain either the observational bias or the dynamical pathway to shepherding $\omega$ that results in the apparent clustering of $\omega$ in the observed sample.

\textbf{6. } 
A massive external peturber generically causes cyclic pericenter oscillations that drive TNOs into Neptune or even Jupiter crossing orbits. 
This process decouples TNOs from any shepherding influence of the external perturber and results in the randomization of $\omega$ and $\Omega$ in the large-$a$ TNOs due to precession.
TNOs interacting with Neptune and the external perturber can become distributed in the large-$a$ region with random $\omega$ and $\Omega$ angles and should be part of the detected sample.
The observed clustering of $\omega$ and $\Omega$ (Figure~\ref{fig:angles}) are not produced in the P9 scenario.

\textbf{7. } 
In exoplanet and debris disk systems with a massive external perturber, $\varpi$ shepherding may have implications for dust production.
The induced shepherding of $\varpi$ aligns orbits in physical space and will bring particles to pericenter in the same angular region, which may enhance the collisional probability and thus dust production for this location.
This may prove an interesting avenue to explore for systems with massive eccentric exoplanets beyond the debris disk and could possibly contribute to a pericenter glow \citep{wyatt99}.

\section{Conclusion}
\label{sec:conclusion}

In this work we have integrated clones of the $a >150$ au, $q > 30$ au TNOs for 4 Gyr in the presence of a candidate P9 perturber, examining the consequences of a distant massive perturber on the TNOs used to infer the planet's existence.
We find that P9 shepherds $\varpi$ for clones with $a > 250$ au, driving this confinement for hundreds of Myr to 4 Gyr.
Clones that experience $\varpi$ shepherding also undergo $q$ oscillations and $i$ flipping, which suggests the presence of a very massive (tens of M$_\Earth$) reservoir of large-$a$ TNOs.
The P9 scenario produces a larger reservoir of potentially detectable yet unseen high-$i$ TNOs with shepherded $\varpi$ values, suggesting there is a currently missing or unseen signature of P9.
The P9 scenario does not produce the observed simultaneous clustering in the angles $\omega$, $\Omega$, and $\varpi$ that is seen in the detected sample.
Taken alone, each of the consequences poses a challenge for the P9 scenario. 
Taken together, these consequences suggest that the existence of a distant massive perturber is unlikely.

A distant massive perturber produces a set of very interesting signatures on a set of large-$a$ planetesimals, but the signature that has driven the newest incarnation of the distant planet hypothesis, namely the clustering of $\omega$, $\Omega$, and $\varpi$, is not produced by this scenario.
Ongoing surveys  (like the Outer Solar Systems Origins Survey \citep{bannister16} and that of \citet{trujillosheppard14,sheppardtrujillo16}) will hopefully provide the detections and proper survey characterizations needed to examine the underlying impetus for the P9 scenario: the apparent clustering in $\omega$, $\Omega$, and $\varpi$ of the large-$a$ TNOs.

\acknowledgments

This project was funded by the National Science and Engineering Research Council and the National Research Council of Canada. This research used the facilities of the Canadian Astronomy Data Centre operated by the National Research Council of Canada with the support of the Canadian Space Agency.
CS gratefully acknowledges support from the NSERC CGS Fellowship.
SML gratefully acknowledges support from the NRC Canada Plaskett Fellowship.



\vspace{5mm}
\facilities{CANFAR, NRC}

\software{Mercury, Python}

\bibliographystyle{apj}
\bibliography{citations.bib}

\allauthors

\listofchanges

\end{document}